\documentstyle[12pt]{article}
\begin{document}
\begin{titlepage}
\begin{flushright}
IASSNS-HEP-95/28
\end{flushright}
\vspace{2.5cm}
\begin{centering}
{\LARGE{\bf Casimir effect around disclinations}}\\
\bigskip\bigskip
\renewcommand{\thefootnote}{\fnsymbol{footnote}}
Fernando Moraes\footnote{On leave from:\\
Departamento de F\'{\i}sica\\
Universidade Federal de Pernambuco\\
50670-901 Recife, PE, Brazil}\\
{\em School of Natural Sciences\\
Institute for Advanced Study\\
Princeton, NJ 08540\\
U.\ S.\ A.}

\end{centering}
\vspace{1.5cm}
\begin{abstract}

This communication concerns the structure of the electromagnetic quantum vacuum
in a disclinated insulator. It is shown that a nonzero vacuum energy density
appears when the rotational symmetry of a continuous insulating elastic medium
is broken by a disclination. An explicit expression is given for this Casimir
energy density in terms of the parameter describing the disclination.

\end{abstract}
\end{titlepage}
\def\carre{\vbox{\hrule\hbox{\vrule\kern 3pt
\vbox{\kern 3pt\kern 3pt}\kern 3pt\vrule}\hrule}}

\baselineskip = 18pt

\section{Introduction}
Space-time being isotropic and homogeneous implies that the vacuum state of a
quantum field must be invariant under translations or rotations. The presence
of external fields, a change in the geometry (curvature, torsion) or a change
in the topology (defects, boundaries) of the space breaks the underlying
symmetry of space-time, inducing changes in the vacuum state of the quantum
field as compared to its field-free, unbound, Minkovski space counterpart. This
is the celebrated Casimir effect~\cite{Mos,Plu}.

Topological defects certainly offer an interesting  (and quite unexplored)
arena for the study of vacuum polarization phenomena in condensed matter
systems. In previous publications~\cite{Fur1,Fur2} we studied the
non-relativistic quantum mechanics of electrons and holes in the presence of a
disclination in an insulating elastic medium, with and without an external
magnetic field. In the present work, I study the Casimir effect of the
electromagnetic field in an infinite medium of constant dielectricity
$\varepsilon$ and permeability $\mu=1$ (a nonmagnetic medium), with a
disclination at the origin of the coordinate system, at zero temperature. It is
investigated the electromagnetic Casimir effect since it is more likely to be
experimentally observable in the solid state than the fermionic
case~\cite{Dur}. This is due to the fact that massive fields, like the fermion
field, give Casimir energies exponentially decreasing with their mass. An
estimate~\cite{Dur} for electrons confined to regions of nanometric s

ize, i.e. $\sim10^{-7}$ cm, gives an exponential damping factor for the Casimir
energy of $\sim5\times10^3$. While the electromagnetic Casimir energy for light
confined to a similar region is $\sim10eV$, the electronic effect is then
roughly $e^{-5\times10^3}$ weaker. Notice that while reference~\cite{Dur} deals
with the Casimir effect due to confinement to mesoscopic regions, in this work
the interest is in the effect outside a defect - no confinement envolved. In
any case, one should expect a much weaker electronic Casimir effect, then
justifying the choice of presenting the electromagnetic effect here.

\section{QED in conical space}
Quantum field theory has been extensively studied in the conical space-time of
cosmic strings~\cite{Gib}. The large body of information available on this
subject makes an excelent guide for the study of quantum field theoretical
aspects of disclinations in solids, in the context of the theory of
defects/three-dimensional gravity of Katanaev and Volovich~\cite{Kat}.

Disclinations, like cosmic strings, are associated with a conical
metric~\cite{Kat} whose spatial part is given, in cylindrical coordinates, by
\begin{equation}
ds^2=dz^2+d\rho^2+\alpha^2\rho^2d\varphi^2.
\end{equation}
With $\alpha=1+\frac{\lambda}{2\pi}$, the disclination is obtained by either
removing ($\lambda<0$) or inserting ($\lambda>0$) a wedge of material, of
dihedral angle $\lambda$, to the otherwise smooth elastic continuum. The
resulting space, described by the above metric, has a null curvature tensor
everywhere, except at the defect where it has a two-dimensional
$\delta$-function singularity given by~\cite{Sok}
\begin{equation}
R^{12}_{12}=R^{1}_{1}=R^{2}_{2}=2\pi\frac{1-\alpha}{\alpha}\delta_{2}(\rho),
\end{equation}
where $\delta_{2}(\rho)$ is the two-dimensional delta function in flat space.
It is clear that $0\leq\alpha<1$ corresponds to a positive-curvature
disclination and $\alpha>1$ to a negative-curvature disclination. $\alpha=1$,
of course, describes the Euclidean medium (absence of disclinations).

The relevant physical quantity in the study of the Casimir effect is the vacuum
expectation value of the energy-momentum tensor, or stress tensor,
$<\hat{T}_{\mu}\,\,^{\nu}>$. Since this tensor diverges, in order to obtain
physically meaningful results, one should subtract from it its also divergent
flat space version\footnote[1]{In many cases an additional regularization is
needed to ensure a finite result.}. That is, one should calculate  the
renormalized vacuum expectation value
\begin{equation}
<\hat{T}_{\mu}\,\,^{\nu}>_{ren} = <\hat{T}_{\mu}\,\,^{\nu}>_{discl} -
<\hat{T}_{\mu}\,\,^{\nu}>_{flat}.
\end{equation}
In this work, I will be concerned only with the $<\hat{T}_{0}\,\,^{0}>$
component, since it is the one that gives the energy density.

Being the electromagnetic field a gauge field, care must be exercised in order
that appropriate gauge-fixing be taken into account in the quantum theory. As
it turns out~\cite{Fro} the convenient gauge here is
\begin{equation}
\nabla_{i}A^{i}=\nabla_{a}A^{a}=0\,\,\,(i=0,1;\,a=2,3),
\end{equation}
where $A^{\mu}$ is the electromagnetic vector potential.
In this gauge the energy density assumes the very simple form
\begin{equation}
<\hat{T}_{0}\,\,^{0}>=i\hbar c\lim_{x' \rightarrow x}
\nabla_{i}\nabla^{i}G(x-x'),
\end{equation}
where $G(x-x')$ is the scalar Green's function, which in the absence of the
defect is
\begin{equation}
G_{0}(x-x') = -\frac{i}{4\pi^{2}\varepsilon}\frac{1}{(x-x')^2}.
\end{equation}
$x-x'$ is the space-time interval between $x$ and $x'$, its square being given
by
\begin{equation}
(x-x')^2 = -c^{2}(t-t')^2 + (z-z')^2 + \rho^2 + \rho '^2 - 2\rho\rho '\cos
(\varphi - \varphi ').
\end{equation}
The energy density is then
\begin{equation}
<\hat{T}_{0}\,\,^{0}>_{ren} =i\hbar c
(-\frac{1}{c^2}\partial_{t}^{2}+\partial_{z}^{2})(G_{\alpha}(x-x')-G_{0}(x-x'))\mid_{x'=x},
\end{equation}
where $G_{\alpha}(x-x')$ is the scalar Green's function in the medium with
boundary conditions. Due to the disclination, the boundary condition in this
problem is a periodicity of $2\pi\alpha$ in the angular variable $\varphi$.
Accordingly, imposition of this condition on the free space Green's function
can be done by reperiodising it via the contour integral~\cite{Dow1}
\begin{equation}
G_{\alpha}(x-x') = \frac{-i}{4\pi\alpha}\int_{A} d\gamma\,
G_{0}(\varphi-\varphi'-\gamma)\,\frac{\exp(i\gamma/2\alpha)}{\sin
(\gamma/2\alpha)}.
\end{equation}
The singularities in the integrand are poles that can be easily found by
writing
\begin{displaymath}
(x-x')^2 = 2\rho\rho' [\cosh\beta-\cos (\varphi-\varphi')],
\end{displaymath}
where
\begin{displaymath}
\cosh\beta = \frac{-c^{2}(t-t')^2+(z-z')^2+\rho^2+\rho '^2}{2\rho\rho'}.
\end{displaymath}
Since $\cosh\beta=\cos i\beta$, the poles of $G_{0}(\varphi-\varphi '- \gamma)$
are at $\gamma=\pm i\beta+\varphi-\varphi'+n2\pi$. On the other hand,
$\sin(\gamma/2\alpha)$ contributes with poles at $\gamma=n2\pi\alpha$. In both
cases $n\in Z$.

As shown in figure 1, the contour $A$ has two branches, one in the upper half
$\gamma$-plane from $(\pi+\varphi-\varphi')+i\infty$ to
$(-\pi+\varphi-\varphi')+i\infty$ passing below the singularity
$\gamma=i\beta+\varphi-\varphi'$ and the other in the lower half-plane from
$(-\pi+\varphi-\varphi')-i\infty$ to $(\pi+\varphi-\varphi')-i\infty$ passing
above the singularity at $\gamma=-i\beta+\varphi-\varphi'$.

In the limit $x'=x$ the poles at  $\gamma=\pm i\beta+\varphi-\varphi'+n2\pi$
will be lying at $\gamma=n2\pi$. So, one is left with poles at $\gamma=n2\pi$
and $\gamma=n2\pi\alpha$. Considering this, there is a deformation~\cite{Dow1}
of the contour $A$ that will prove to be quite convenient. Just deform it into
an anticlockwise loop $B$ around the pole at $\gamma=0$ and the two vertical
lines, $\Gamma$, $\gamma=b+iy$ and $\gamma=-b+iy$, with $-\infty<y<\infty$ and
$b$ a constant (see figures 2 and 3). This can only be done if the poles lying
on the Re$\gamma$-axis, other than the one at $\gamma=0$, are outside the
region limited by the lines; i.e., $b$ must be chosen to be less than the
smaller of $2\pi$ and $2\pi\alpha$. As shown in the appendix, the contribution
from loop $B$ turns out to be exactly $G_{0}(x-x')$. Since renormalization
requires  $G_{\alpha}(x-x')-G_{0}(x-x')$, all one has to consider is the
contribution from the two vertical lines $\Gamma$.
Equation (8) becomes then
\begin{equation}
<\hat{T}_{0}^{\,\,0}>_{ren} =  \frac{-i\hbar c}{4\pi\alpha}\int_{\Gamma}
d\gamma \frac{\exp(i\gamma/2\alpha)}{\sin (\gamma/2\alpha)}
\,(-\frac{1}{c^{2}}\partial_{t}^{2}+\partial_{z}^{2})
G_{0}(\varphi-\varphi'-\gamma)\,\mid_{x'=x}.
\end{equation}

It is easily found that
\begin{equation}
(-\frac{1}{c^{2}}\partial_{t}^{2}+\partial_{z}^{2})
G_{0}(\varphi-\varphi'-\gamma)\,\mid_{x'=x} =
\frac{i}{16\pi^{2}\varepsilon\rho^{4}\sin^{4}(\gamma/2)}.
\end{equation}
Furthermore, using the symmetry of the contour $\Gamma$ to replace the
exponential by a cosine in equation (10) and then deforming the contour into a
clockwise loop around the origin leads to
\begin{equation}
<\hat{T}_{0}\,\,^{0}>=\frac{i\hbar c}{64\pi^{3}\varepsilon\rho^{4}\alpha} \oint
d\gamma \frac{\cot(\gamma/2\alpha)}{\sin^{4}(\gamma/2)}.
\end{equation}
This integral can be easily evaluated using calculus of residues, by expanding
the integrand in powers of $\gamma$  such that the term of order $\gamma^{-1}$
gives $\oint\frac{d\gamma}{\gamma}=-2\pi i$. The expansion results in
\begin{equation}
\frac{\cot(\gamma/2\alpha)}{\sin^{4}(\gamma/2)}=32\alpha\gamma^{-5} +
(\frac{16}{3}\alpha - \frac{8}{3\alpha})\gamma^{-3} +
(\frac{22}{45}\alpha-\frac{4}{9\alpha}-\frac{2}{45\alpha^{3}})\gamma^{-1}+O(\gamma^{0}).
\end{equation}
The energy density is then
\begin{equation}
<\hat{T}_{0}\,\,^{0}>=\frac{\hbar c}{720\pi^{2}\varepsilon\rho^{4}}
(11-\frac{10}{\alpha^2}-\frac{1}{\alpha^4}).
\end{equation}
This result has been found by a variety of ways in the context of cosmic
strings~\cite{Fro,Bro,Dow2} and also for a metallic wedge~\cite{Deu}.

Notice that in the flat space limit, $\alpha=1$,
$<\hat{T}_{0}\,\,^{0}>_{ren}=0$ as it should. Also, although values of $\alpha
>1$ are probably not physically sensible for cosmic strings because it would
correspond to negative linear mass densities, in condensed matter it describes
negative-curvature disclinations~\cite{Fur1,Fur2}. The immediate consequence
for  condensed matter systems with such defects is a dependence of the sign of
the force exerted on the defect by the electromagnetic vacuum, on the sign of
the defect curvature. This is certainly important for the study of the
stability of the defects. The obvious implication is that a disclination
dipole, or edge dislocation~\cite{Kat}, is more stable than either disclination
standing by itself and that the interaction between the pair is attractive.
This is a result of the vacuum polarization of the electromagnetic field only.
The total interaction and overall stability depend on contributions from all
the relevant fields.

\section{Concluding remarks}
The purpose of this Letter is to draw attention to the importance of quantum
field theoretical effects in condensed matter enclosing topological defects. A
very simple case has been considered: the medium is infinite (e.g. no external
boundaries were considered), nonmagnetic, has only one defect (disclination),
its dielectricity is frequency-independent and the temperature set to $T=0$. Of
course, these oversimplifications give a distorted picture of the real
phenomenum. Nevertheless, it is shown that a finite electromagnetic Casimir
energy density  is present in the outer region of the disclination.
It would be interesting to extend the present results to finite temperatures
and to magnetic media. Moreover, use of a frequency-dependent complex
dielectric constant would not only provide a more realistic calculation for the
Casimir energy but also one that takes into account absorption of energy inside
the dielectric, something that happens in real materials.
\\
\noindent
{\bf Acknowledgements}\\
\noindent
This work was partially supported by CNPq.
\\
\begin{center}
{\bf APPENDIX}\\
\end{center}
\noindent
The anticlockwise loop integral
\begin{displaymath}
\frac{-i}{4\pi\alpha}\oint_{B}d\gamma
\frac{\exp(i\gamma/2\alpha)}{\sin(\gamma/2\alpha)}
G_{0}(\varphi-\varphi'-\gamma)
\end{displaymath}
can be evaluated in the same way as the integral in equation (12). Again,
symmetry of the contour permits the substitution of the exponential by a
cosine. The resulting $\cot(\gamma/2\alpha)$ expanded around $\gamma=0$ is
\begin{displaymath}
\frac{2\alpha}{\gamma}-\frac{\gamma}{6\alpha}-\frac{\gamma^{3}}{360\alpha^{3}}+O(\gamma^{5}),
\end{displaymath}
implying
\begin{displaymath}
\frac{-i}{4\pi\alpha}\oint_{B}d\gamma
\frac{\exp(i\gamma/2\alpha)}{\sin(\gamma/2\alpha)}
G_{0}(\varphi-\varphi'-\gamma)=\frac{-i}{2\pi}\oint_{B}\frac{d\gamma}{\gamma}
G_{0}(\varphi-\varphi'-\gamma)=G_{0}(x-x').
\end{displaymath}

\noindent
{\Large {\bf Figure Captions}}
\\
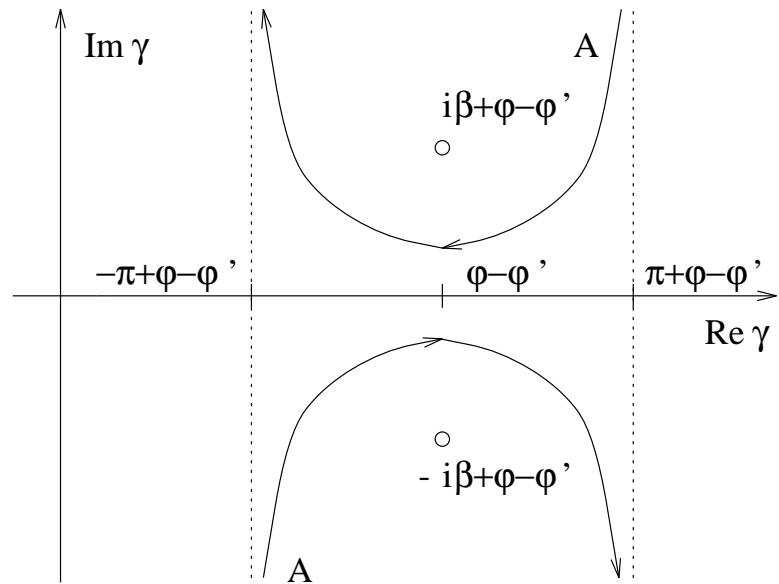
Figure 1: Integration contour for equation (9).
\\
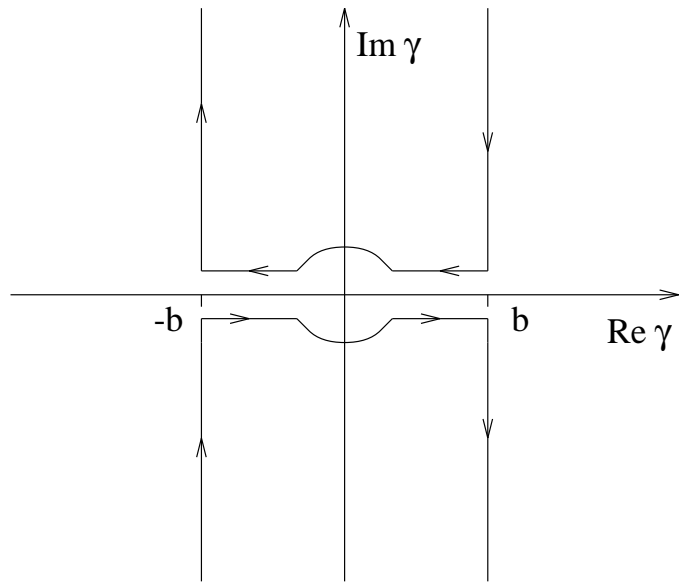
Figure 2: Intermediate deformation of the contour.
\\
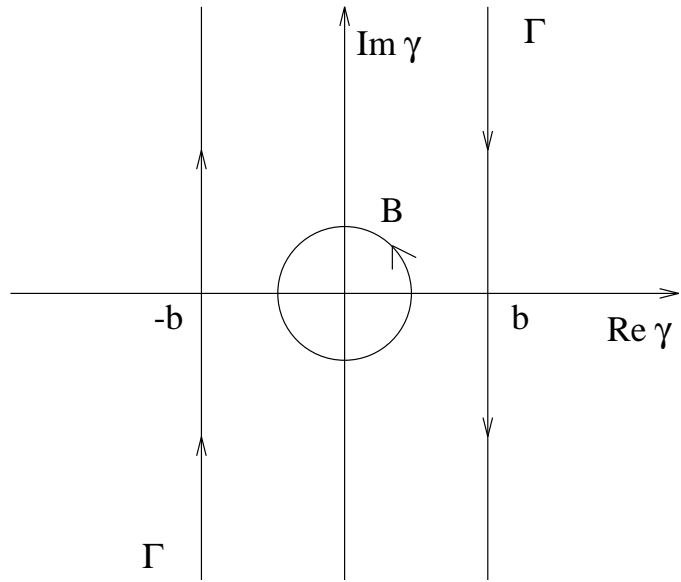
Figure 3: Final form of the deformed contour.
\end{document}